\documentstyle[11pt,newpasp,twoside,epsf]{article}
\def\edcomment#1{\iffalse\marginpar{\raggedright\sl#1\/}\else\relax\fi}
\reversemarginpar

\begin{document}
\title{Evolution of D and $^3$He in the Galaxy}
 \author{Monica Tosi}
\affil{Osservatorio Astronomico di Bologna, Via Ranzani 1, I-40127 Bologna,
Italy}

\begin{abstract}
The predictions of Galactic chemical evolution models for D and $^3$He are 
described in connection with those on the other Galactic quantities
for which observational constraints are available.

Models in agreement with the largest set of data predict deuterium depletions 
from the Big Bang to the present epoch smaller than a
factor of 3 and do not allow for D/H primordial abundances larger 
than $\sim4\times10^{-5}$.  Models predicting higher D consumption do 
not reproduce other observed features of our Galaxy.
 
If both the primordial D and $^3$He are low, models assuming that 90$\%$ of 
low-mass stars experience an extra-mixing
during the red giant phase reproduce all the $^3$He observed abundances.
The same percentage allows to fit also the observed carbon isotopic ratios,
thus supporting the self-consistency of the extra-mixing mechanism.

\end{abstract}

\section{Introduction}
In this review, I will try to describe what Galactic chemical evolution
models tell us about the evolution and the primordial abundances of D 
and $^3$He, and what, in turn, D and $^3$He may tell
us about stellar and Galactic evolution. In particular, I wish
to emphasize that the light elements should not be treated separately, but 
should always be considered together with the other more diffuse
elements, to better constrain their evolution.

The reason why Galactic chemical evolution models are required to derive
the primordial D and $^3$He abundances from the observed ones is that
all the objects where the two elements are measurable are relatively young 
(the oldest being the sun with an age of 4.5 Gyr) and have therefore 
formed, with the only exception of high-redshift clouds,  
from an ISM whose chemical composition had 
been modified by the previous stellar generations. To infer the primordial 
abundances from these measurements, it is thus necessary to take into 
account the effects of the various cycles of gas astration and gas return, 
and the variations of the ISM chemical composition due to stellar 
nucleosynthesis and gas flows occurring up to the time when the 
observed objects have formed. This is accomplished by chemical 
evolution models. 

D and $^3$He are obviously related to each other, since all the D which
enters a star is immediately burnt into $^3$He (Reeves et al. 1973). 
However, the problems faced when studying their Galactic evolution are quite 
different, and I will thus treat them separately in this paper.

\section{D evolution}

Since D is completely destroyed inside stars already in pre-main sequence
phase, if we consider the Big Bang nucleosynthesis as the only source of
D, the amount of D which can be present in any galactic region and at
any place is that contained in gas which has never been through stars.
In other words, the fraction of primordial D surviving at any epoch and
in any region is equal to the fraction of virgin gas there. Hence, in
principle, to infer the primordial abundance of D from its present one 
it would be sufficient to know the current fraction of gas which has 
not entered a star yet (Steigman \& Tosi 1995). Unfortunately, we do not 
know the fraction
of pristine gas even in the most local medium and we must therefore rely
on Galactic chemical evolution models to derive the D evolution.

\begin{figure}
\vspace{8.5truecm}
\includegraphics{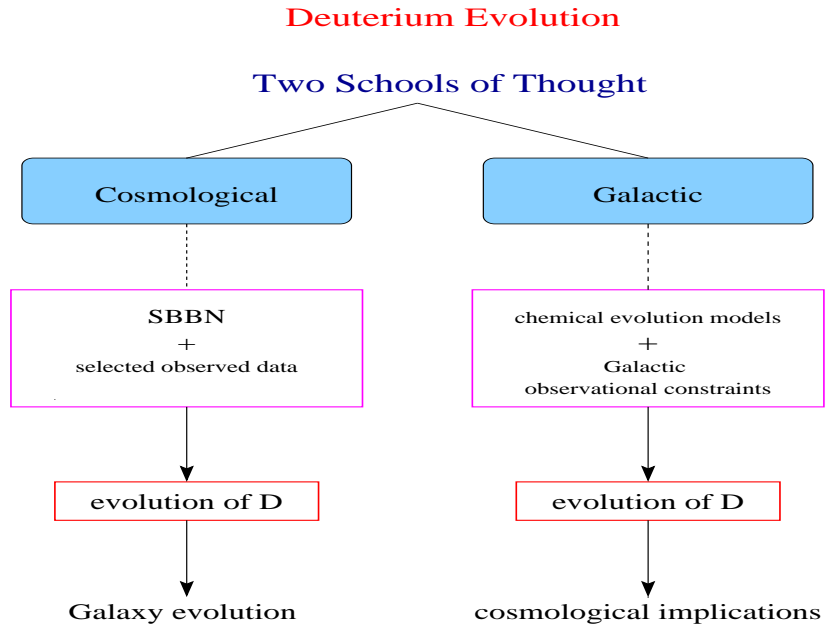}
\caption{Sketch of the two main approaches to study the evolution of D.}
\end{figure}

Historically, there are two schools of thought on how to proceed in
studying the D evolution, as sketched in Fig.1. The first one, 
in chronological order, can be referred to as the {\it Cosmological School}.
The approach of this school is to start from standard Big Bang 
nucleosynthesis (SBBN) prescriptions, select the observational constraints on 
D which can be considered reliable, and infer from these sets of data 
what the D evolution must have been in the Galaxy. They then build models
of Galaxy evolution able to reproduce the inferred trend of D vs time,
and the predictions of such models on the other Galactic quantities
are a by-product.

The other school, which I will call {\it Galactic}, follows the opposite
approach. We start from chemical evolution models of our Galaxy, select
only those which are able to reproduce the largest set of observational 
constraints, and take the predictions on D only from these selected models.
The consequences of these predictions on the primordial D abundance and on
cosmology are a by-product. 
Were we living in the best of all possible worlds, the two approaches
should provide the same results. Instead, their predictions 
are quite different from each other.

Fig.2 shows the D abundances derived from all the available
observations and plotted as a function of the supposed epoch of formation
of the observed objects. All the error bars are $2\sigma$. 
The two vertical bars at t=0 represent the ranges of values derived by 
Songaila et al. (1994, hereinafter SCHR94) and Burles \& Tytler 
(1998, B\&T98) from high-redshift, low-metallicity, absorbers
on the line of sight of distant QSO's. The bar at 8.5 Gyr represents the
value of the Protosolar Cloud inferred by Geiss \& Gloeckler (1998, G\&G98) 
from solar system data. The solid bar at 13 Gyr shows the range of abundances
derived by  Linsky (1998, L98) for the local ISM, while
the length of the dotted bar shows the possible cloud-to-cloud variations
suggested by Vidal-Madjar, Ferlet \& Lemoine (1998, V-M98). 

\begin{figure}
\vspace{6.8truecm}
\includegraphics{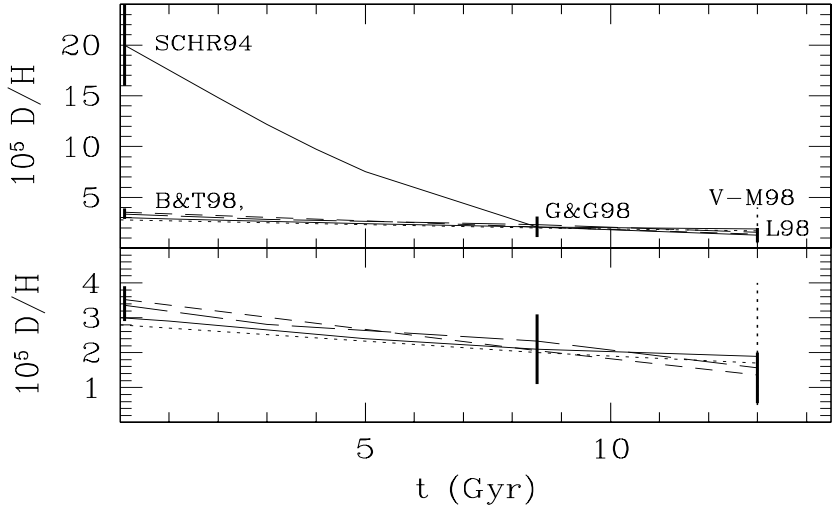}
\caption{Top panel: Observational abundances of D as a function of the 
target formation epoch. The steep curve sketches the local evolution of 
D as proposed by the {\it Cosmological School}, the other curves show 
the predictions of chemical evolution models in
agreement with the largest set of observed Galaxy properties. 
Bottom panel: blow-up of the lower part of the top panel.
See text for details. 
}
\end{figure}

Since the {\it Cosmological School} was founded when the primordial $^4$He 
was definitely supposed to be low (Y$_P\simeq$0.23 in mass fraction), and 
since SBBN predicts the primordial D to be anti-correlated with Y$_P$, members
of this school obviously thought that the only reliable measures of D in
almost primeval systems, like high-redshift, low-metallicity absorbers,
were those leading to high D abundances.
They thus thought that the natural evolution of deuterium with time is that
connecting the SCHR94 value with the local ones and sketched by the solid
line in Fig.2, i.e. a D destruction by one order of magnitude
from the primordial to the present abundance.

To obtain such a high D destruction during the Galaxy evolution, one must
invoke a high star formation rate (SFR), which is usually assumed to
occur at the earliest epochs, because all the observational evidences are
against high SFR at relatively recent times. These high SFRs (and their related
metal enrichment), inevitably imply a large overproduction of the heavy
elements with respect to the observed stellar abundances, unless compensated
by mechanisms able to reduce the excess of metals, by diluting 
or removing them from the Galaxy. For this reason, models with high 
D destruction usually invoke infall of metal poor gas and galactic
winds powered by supernovae explosions, sometimes coupled with variations
in the initial mass function. There have been several attempts to find
viable Galactic models with strong deuterium depletion, but no scenario 
consistent with all the Galactic data has been found. For instance, in their 
pioneering work, Vangioni-Flam \& Audouze (1988) concluded that they 
excessively overproduced the metals, and Scully et al. (1997), in order to 
obtain the desired D without overproducing the metals, ended up with a present
local SFR at least one order of magnitude lower than observed.
Tosi et al. (1998) have tested all the possible combinations of the various
parameters (SFR, infall, winds, etc.) and have always found significant
inconsistencies in the models with high D destruction: metal overabundance  
with wrong galactocentric distribution, or metallicity distribution of 
the G-Dwarfs in the solar neighbourhood completely at odds
with the observed one, or abundance ratios in halo or disk stars different
from the observed ones (e.g. [O/Fe] vs [Fe/H]), or SFR inconsistent with
the observed range. In no way have we been able to find a fairly
self-consistent model with high D destruction.

The {\it Galactic School} works instead on chemical evolution models able
to reproduce as well as possible the largest set of observed Galactic
features. 
Thanks to the improvements both on the observational and on the theoretical
sides, good chemical evolution models of the Milky Way nowadays can reproduce
the average distribution of the following list of observed features
(see e.g. Tosi 1996 and 2000, Boissier \& Prantzos 1999 for references):
\par\noindent
$\bullet$ current distribution with Galactocentric distance of the SFR,
\par\noindent
$\bullet$ current distribution with Galactocentric distance of the gas 
 and star densities,
\par\noindent
$\bullet$ current distribution with Galactocentric distance of element 
 abundances as derived from HII regions and from B-stars, 
\par\noindent
$\bullet$ distribution with Galactocentric distance of element abundances at
 slightly older epochs, as derived from PNe II, 
\par\noindent
$\bullet$ age-metallicity relation not only in the solar neighbourhood but also
 at other distances from the center,
\par\noindent
$\bullet$ metallicity distribution of G-dwarfs in the solar neighbourhood, 
\par\noindent
$\bullet$ local Present-Day-Mass-Function (PDMF),
\par\noindent
$\bullet$ relative abundance ratios (e.g. [O/Fe] vs [Fe/H]) in disk and halo
 stars. 

When one compares with each other all the models
in better agreement with these data (e.g. Tosi 1996), the striking result
is that they all predict essentially the same deuterium evolution, in spite
of the fact that they are computed by different people, with different
assumptions on the input parameters and with different numerical codes.
The bottom panel of Fig.2 shows an updated version of the comparison:
the plotted models are from Galli et al. (1995a, short-dashed line),
Dearborn, Steigman, \& Tosi (1996, solid line), Chiappini \& Matteucci 
(1996, long-dashed line) and Boissier \& Prantzos (1999, dotted line).
All the shown curves fit very well the average abundances
derived for the local ISM, the pre-solar nebula and the high-redshift
absorbers by B\&T98. They all show a fairly moderate
(a factor from 1.5 to 3, at most) D destruction during the Galaxy 
lifetime, and therefore suggest that the primordial D abundance should
be low: 2$ \leq $(D/H)$_P\times 10^5 \leq 4$.

This homogeneity of predictions is not a chance effect, but
the consequence of the circumstance that all these models fit equally
well the observational data on the present SFR, gas and mass densities,
and chemical abundances, which necessarily implies that they predict
similar fractions of pristine gas and, therefore, of surviving primordial
deuterium.

Our current knowledge on the Galactic evolution of D can thus be
summarized as follows: Models predicting high deuterium destruction
cannot account for all the observed Galactic
properties; models able to reproduce the largest set of Galactic
properties all predict low deuterium destruction and, hence, low
primordial D.

\section {$^3$He evolution}

\begin{figure}
\vspace{5.2truecm}
\includegraphics{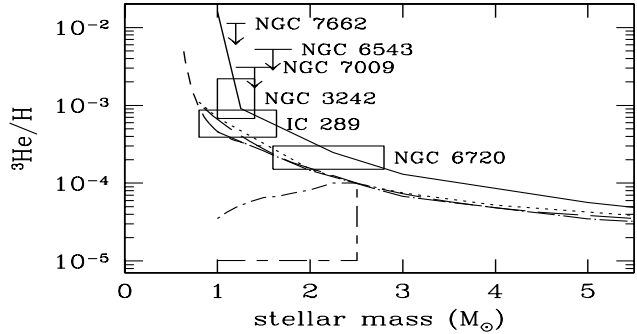}
\caption{$^3$He yield as a function of  
the stellar initial mass. The curves show various stellar nucleosynthesis
predictions, the boxes and arrows the PNe with high $^3$He (see Galli 
et al. 1997 for details).
}
\end{figure}

$^3$He has a more complex evolution than D, because it is produced not only
during the Big Bang but also inside stars, during the
main sequence phase. This early stellar production may be however 
largely compensated by further nuclear processing in subsequent phases.
Standard stellar nucleosynthesis studies predict that, at the end of the
star life, the $^3$He present in the initial stellar composition is
significantly destroyed in massive stars, but preserved or even strongly
enhanced in lower mass stars, and that the  $^3$He net yield 
is a steeply decreasing function of the stellar initial mass, with a
large net production in stars below 2--2.5 M$_{\odot}$ (see e.g the
monothonic curves in Fig.3). This behaviour was known since
the late sixties (e.g. Iben 1967), and already in 1976 Rood, Steigman,
\& Tinsley noticed that it leads to overproduce the solar 
abundance. Only in the mid-nineties, however, with the advent of more
detailed  combinations of $^3$He yields with galactic chemical evolution 
models  (Vangioni-Flam, Olive \& Prantzos 1994, Galli et al. 1995a, 
Dearborn, Steigman
\& Tosi 1996) it became apparent that the results on  $^3$He of standard 
nucleosynthesis studies are definitely inconsistent both
with the solar and with the ISM observed abundances. This inconsistency
is found with any type of Galactic evolution models, including 
those in agreement with all the other observational constraints
(e.g. Tosi 1996) and was emphasized by several groups at the Elba
meeting on the Light Element Abundances in 1994 (Cass\'e \& Vangioni-Flam
1995, Galli et al. 1995b, Tosi, Steigman \& Dearborn 1995). In that occasion,
Michel Cass\'e concluded with what has been the most popular refrain on
$^3$He ever since: {\it $^3$He delendum est}, like the city of Carthago
for the ancient Roman M.P. Cato Censor.

The most probable solution to the  $^3$He problem is less drastic than
that applied to Carthago by the Romans and was proposed already
in 1995 (Charbonnel 1995, Hogan 1995). It consists in the further $^3$He
processing into heavier elements favoured by an extra-mixing occurring 
in the red giant phase of low-mass stars (see both Charbonnel and Sackman, 
this volume). When low-mass stars are assumed to experience this
extra-mixing and the so-called Cool Bottom Processing (CBP), Galactic 
evolution models do not overproduce  $^3$He
anymore and fit well the observed solar and HII region abundances
(Tosi 1996). 
The question is: in what fraction of low-mass stars CBP should occur to 
best fit all the data, taking into account that Bania, Rood et al. 
(this volume)  measure in a few PNe a high  $^3$He perfectly consistent 
with the predictions of standard stellar nucleosynthesis (Fig.3) ? 
Galli et al. (1997) showed that the fraction should be larger than
80$\%$ to fit the $^3$He abundances observed in the solar system 
(Geiss \& Gloeckler 1998), in PNe and in HII regions (Rood et al. 1995 and 
this volume). 

\begin{figure}
\vspace{4.3truecm}
\includegraphics{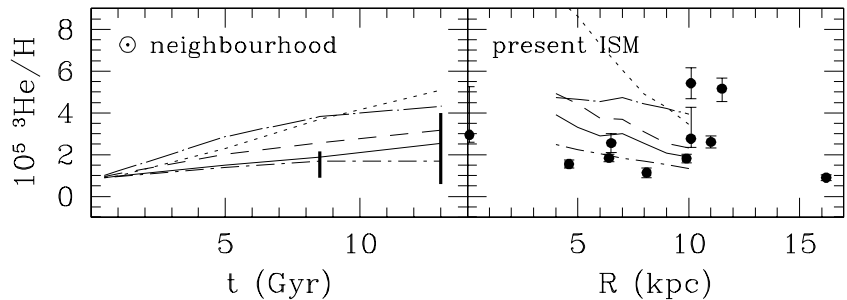}
\caption{Comparison between the observed abundances of $^3$He and the 
predictions of model Tosi-1 with CBP in 0, 90 or 100\% of low-mass stars. 
Left hand panel: Evolution of $^3$He/H in the solar neighbourhood. 
Right hand panel: Corresponding radial distribution at the present 
epoch. See text for symbols and references on the observational data.
}
\end{figure}

Fig.4 shows the predictions of the best of models Tosi-1 (see Tosi 1988, 
Dearborn et al. 1996) when 0\% (dotted line in both panels), 90$\%$ (solid 
lines), or 100$\%$ (short-dash-dotted lines) of stars with
M $\leq$ 2.5~M$_{\odot}$ are assumed to follow Sackman \& Boothroyd's
(1999) prescriptions for CBP $^3$He depletion. For the remaining low-mass
stars, as well as for all the intermediate and high-mass ones, the $^3$He yield
is taken from Dearborn et al. (1996). The dotted, solid and short-dash-dotted 
lines correspond to models
assuming as initial abundances (D/H)$_p=3 \times 10^{-5}$ and 
($^3$He/H)$_p=1 \times 10^{-5}$. The dashed lines show the predictions of
the same model with 90$\%$ CBP, when only the initial D is changed to 
(D/H)$_p=10 \times 10^{-5}$, while the long-dash-dotted lines correspond to
(D/H)$_p=20 \times 10^{-5}$.

The vertical bars in the left hand panel represent the ranges of $^3$He
abundances (at 2$\sigma$) derived by Geiss \& Gloeckler (1998) and
Gloeckler \& Geiss (1998) for the Protosolar and the Local Interstellar
Clouds, here assumed to be representative of the local ISM, 4.5 Gyr ago
and now, respectively. The data points in the right hand panel show the
$^3$He abundances derived by Rood et al. (1995) from HII region radio
observations. It is apparent that the models assuming 90$\%$ and 100$\%$ 
of low-mass stars with CBP fit quite well all the data when the initial
D is sufficiently low. The CBP depletion is however insufficient
to compensate the $^3$He overproduction if the initial D/H, subsequently
turned into $^3$He, is higher than a few 10$^{-5}$, in which case, first
the observed protosolar abundance, and then also the local ISM one, cannot
be reproduced any more. This is a further argument in favour of the low
primordial deuterium resulting from the previous section and from Tytler's
(this volume) discussion of the observations at high redshift.

Hence, if (D/H)$_p \simeq 3 \times 10^{-5}$, the $^3$He problem is solved if
90\% of low-mass stars burn it during the extra-mixing occurring in
their red giant phase. In fact, we can simultaneously reproduce the low 
$^3$He abundances of the solar region and of HII regions at any 
Galactocentric distance, and the high abundance of NGC 3252 and the other PNe
measured by Rood et al., which would consequently be associated to the
remaining fraction (10\%) of stars without deep mixing. 

Since the deep mixing depletes not only $^3$He, but also the $^{12}$C/$^{13}$C 
ratio (see Charbonnel and Sackman, this volume), it is important to
check the self-consistency of the solution by comparing the model
predictions with the carbon isotopic ratio. 
Charbonnel and do Nascimento (1998) find indeed that more than 90\% of 191
field and cluster red giants present carbon ratios significantly lower 
than the $^{12}$C/$^{13}$C=25 predicted by standard nucleosynthesis. What
we also want to check are the predictions of chemical evolution models.
This has been done by
Palla et al. (2000, hereinafter PBSTG) with a two-folding approach: a) we
have compared the available observational data on the carbon isotopic ratio
with the corresponding predictions of chemical evolution models assuming the
deep mixing in various percentages of low-mass stars; b) we have observed
$^{12}$C and $^{13}$C in 28 PNe in mm-waves and compared the
derived ratios with those predicted by stellar nucleosynthesis.

\begin{figure}
\vspace{4.7truecm}
\includegraphics{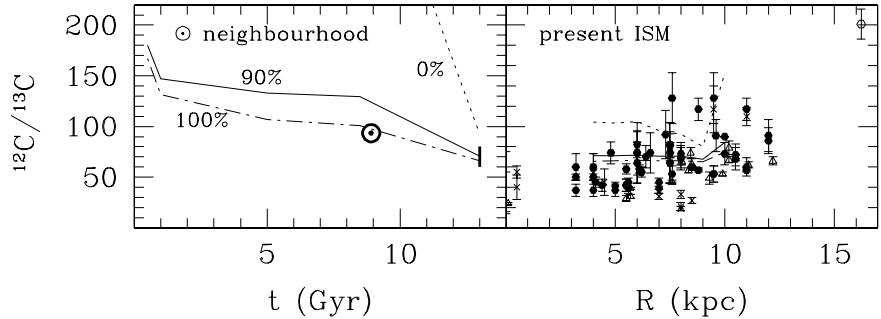}
\caption{Comparison between observed carbon isotopic
ratios and the predictions of model Tosi-1 with CBP in the indicated
fraction of low-mass stars (PBSTG). The data refer to the
sun and to molecular clouds in the Galactic disk (see PBSTG for
references).
Left hand panel: Evolution in the solar neighbourhood. 
Right hand panel: Corresponding radial distribution at the present 
epoch. 
}
\end{figure}

Fig.5 shows what model Tosi-1 predicts for the carbon ratio when the 
$^{12}$C and $^{13}$C adopted yields are from Boothroyd \& Sackman (1999) 
for low-mass stars with CBP, from Marigo (2000) for low and 
intermediate-mass stars without CBP, and from Limongi, Chieffi \& 
Straniero (2000) for massive stars. Equivalent results are described by
PBSTG for stellar yields from other sources. The dotted line shows that
without extra-mixing in low-mass stars the $^{12}$C/$^{13}$C ratio
is overpredicted with respect to both the abundances observed in the sun
and in molecular clouds (assumed to be representative of the present disk
abundances). Vice versa a good agreement is achieved if the
fraction of stars with CBP is as high as possible (recall that one cannot
assume 100\% because of the few PNe with high $^{3}$He). As discussed by
PBSTG, the amount of predicted $^{12}$C and $^{13}$C strongly depends not 
only on the extra-mixing assumptions but also (mostly) on the assumptions for
the nucleosynthesis in intermediate-mass stars, which are the major 
contributors to the ISM enrichment of the carbon isotopes. However, we 
can safely conclude that the observed carbon ratios are always better 
reproduced by models adopting high percentages of low-mass stars with CBP.

\begin{figure}
\plottwo{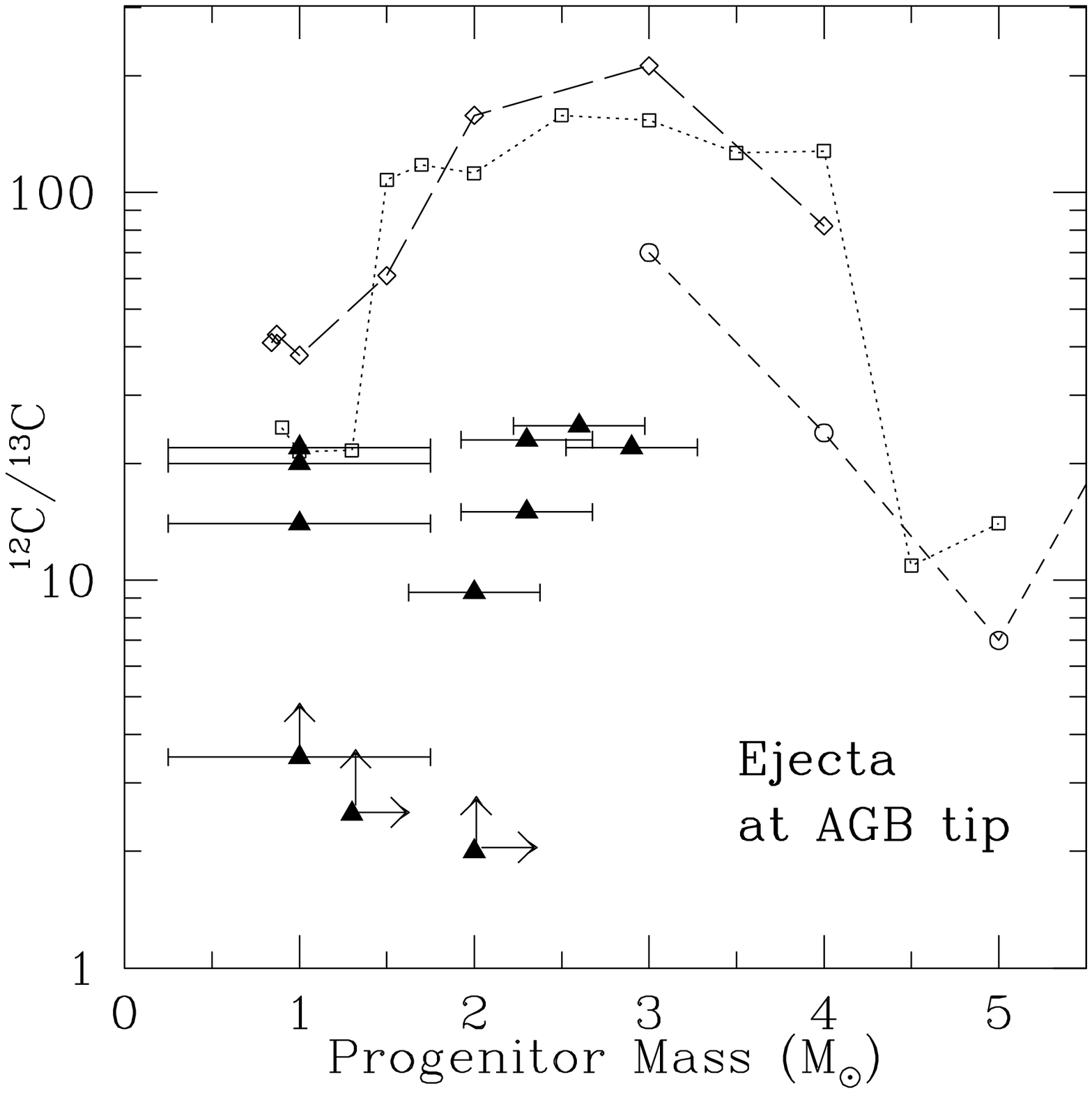}{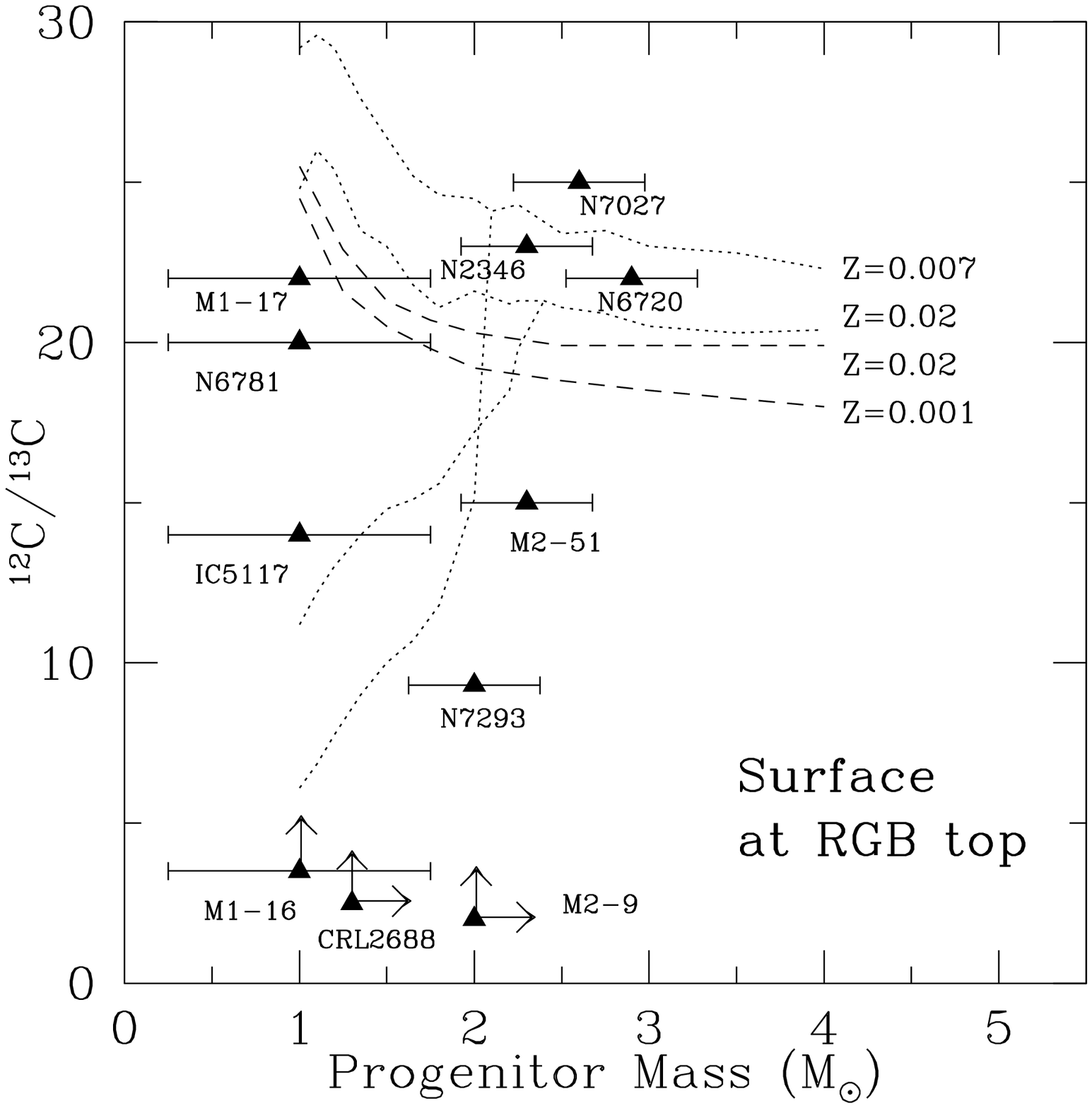} 
\caption{Comparison between the carbon isotopic ratio measured by PBSTG
in PNe and the predictions from stellar nucleosynthesis. Left hand panel: 
composition just before the PN ejection predicted without deep mixing by
Forestini \& Charbonnel (1997, short-dashed curve), van den Hoek \& 
Groenewegen (1997, dotted curve), and Marigo (2000, long-dashed curve).
Right hand panel: composition predicted with and without deep mixing at
the end of the red giant phase (the dotted lines refer to Boothroyd \&
Sackman 1999, the short-dashed ones to Charbonnel 1994).
}
\end{figure}

This is also supported by the comparison of the $^{12}$C/$^{13}$C derived
by PBSTG for the PNe where $^{13}$C was actually measurable
with the carbon ratio predicted for stars right before the ejection of the
PN by various nucleosynthesis studies. The left panel of Fig.6
shows that most of the data points (triangles with associated error on the
progenitor mass estimate) present carbon ratios lower than those expected
from standard nucleosynthesis. The right hand panel shows that the 
measured carbon ratios are consistent with the predictions of the CBP models 
at the end of the red giant phase (unfortunately, no nucleosynthesis models
are available yet up to the pre-PN phase, and one cannot perform a more
appropriate comparison with the PNe observed ratios).

Hence, with deep mixing in $\sim$90\% of low-mass stars one can reproduce 
the abundances of $^{3}$He observed in the sun, in the ISM and in PNe 
and the $^{12}$C/$^{13}$C measured in the sun, in red giants, in the ISM 
and in PNe. We can then conclude that this mechanism appears to be a very
promising process, which needs to be further investigated, both to individuate
its possible causes and to check its effects of later stellar evolution
phases.

Our current knowledge on the Galactic evolution of $^{3}$He can be
summarized as follows: All its available observational abundances can
be explained if a) its primordial abundance is low, ($^{3}$He/H)$_p\simeq1 
\times 10^{-5}$, b) the deterium primordial abundance is also low, and c)
deep mixing occurs in almost all low-mass stars.

\acknowledgements
Most of what has been described in this review results from the collaborations
with C. Chiappini, D. Galli, F. Matteucci, F. Palla, L. Stanghellini and 
G. Steigman, and I thank them all for their help. Interesting 
conversations with C. Charbonnel and N. Prantzos have also been very useful.
I thank the organizers for such an enjoyable and successful symposium.
This work has been partially supported by the Italian COFIN98-MURST at
Arcetri.


\begin{references}

\reference Boissier, S., \& Prantzos, N. 1999, \mnras, 307,  857
\reference Boothroyd, A.I., \& Sackman, I.-J. 1999, \apj, 510, 232
\reference Burles, S., \& Tytler, D. 1998, in Primordial Nuclei and their
 Galactic evolution, N.Prantzos, M.Tosi, \& R. von Steiger eds, \ssr 84, 65,
  B\&T98
\reference Cass\'e, M., \& Vangioni-Flam, E., 1995, in The Light Element
 Abundances, P.Crane ed. (Springer, De), p.44
\reference Charbonnel, C. 1994, \aap, 282, 811
\reference Charbonnel, C. 1995, \apj, 453, L41
\reference Charbonnel, C., \& do Nascimento, J.D.Jr 1998, \aap, 336, 915
\reference Chiappini, C., \& Matteucci, F. 1996, in From Stars to Galaxies: 
 The Impact of Stellar Physics on Galaxy Evolution, C. Leitherer, 
 U. Fritze-von-Alvensleben, \& J. Huchra eds, \pasp Conf.Ser., 98, 541
\reference Dearborn, D.S.P., Steigman, G., \& Tosi, M. 1996, \apj 465, 
 887 (DST96)
\reference Forestini, M., \& Charbonnel, C. 1997, \aaps, 123, 241
\reference Galli, D., Palla, F., Ferrini, F., \& Penco, U. 1995a \apj, 443, 536
\reference Galli, D., Palla, F., Ferrini, F., \& Straniero, O. 1995b, in The 
 Light Element Abundances, P.Crane ed. (Springer, De), p.224
\reference Galli, D., Stanghellini, L., Tosi, M., \& Palla F. 1997 \apj, 477, 
  218
\reference Geiss, J., \& Gloeckler, G. 1998, in Primordial Nuclei and their
 Galactic evolution, N.Prantzos, M.Tosi, \& R. von Steiger eds, \ssr 84, 239,
 G\&G98
\reference Hogan, C.J. 1995, \apj, 441, L17
\reference Iben, I. 1967, \apj,  147, 650
\reference Limongi, M., Chieffi, A., Straniero, O. 2000, in The chemical 
 evolution of the Milky Way: stars versus clusters, F. Matteucci and F. 
 Giovannelli eds (Kluwer, Holland) in press
\reference Linsky, J.L. 1998, in Primordial Nuclei and their
 Galactic evolution, N.Prantzos, M.Tosi, \& R. von Steiger eds, \ssr 84, 285
 L98
\reference Marigo, P. 2000, in The chemical evolution of the Milky Way: 
 stars versus clusters, F. Giovannelli and F. Matteucci eds (Kluwer, Holland)
 in press, astro-ph/9912341 
\reference Palla, F., Bachiller, R., Stanghellini, L., Tosi, M., \& Galli, D.
 2000, \aap, in press, astro-ph/9912086, PBSTG
\reference Reeves, H., Audouze, J., Fowler, W.A., \& Schramm, D.N. 1973, \apj, 
 179, 979
\reference Rood, R.T., Bania, T.M., Wilson, T.L., \& Balser, D.S. 1995, in The 
 light  elements abundances, P.Crane ed. (Springer, De), p.201
\reference Rood, R.T., Steigman, G., \& Tinsley, B.M. 1976, \apj, 207, L57
\reference Sackman, I.-J., \& Boothroyd, A.I. 1999, \apj, 510, 232
\reference Scully, S., Cass\'e, M., Olive, K.A., \& Vangioni-Flam, E. 1997, 
 \apj, 476, 521
\reference Songaila, A., Cowie, L.L., Hogan, C.J., Rugers, M. 1994, Nature,
 368, 599, SCHR94
\reference Steigman, G., \& Tosi, M. 1995, \apj, 453, 173
\reference Tosi, M. 1988, \aap, 197, 33
\reference Tosi, M. 1996, in From Stars to Galaxies: The Impact of Stellar 
 Physics on Galaxy Evolution, C. Leitherer, U. Fritze-von-Alvensleben, \&
 J. Huchra eds, \pasp Conf.Ser. 98, 299
\reference Tosi, M. 2000, in The chemical evolution of the Milky Way: 
 stars versus clusters, F. Giovannelli and F. Matteucci eds (Kluwer, Holland)
 in press, astro-ph/9912370
\reference Tosi, M., Steigman, G., \& Dearborn, D.S.P. 1995, in The Light 
 Element Abundances, P.Crane ed. (Springer, De), p.228
\reference Tosi, M., Steigman, G., Matteucci, F., \& Chiappini, C. 1998, \apj,
 498, 226
\reference van den Hoek, L.B., \& Groenewegen, M.A.T. 1997, \aaps, 123. 305
\reference Vangioni-Flam, E., \& Audouze, J. 1988, \aap, 193, 81
\reference Vangioni-Flam, E., Olive, K.A., \& Prantzos, N. 1994, \apj, 148, 3
\reference Vidal-Madjar, A., Ferlet, R., Lemoine, M. 1998, in Primordial 
 Nuclei and their Galactic evolution, N.Prantzos, M.Tosi, R. von Steiger eds, 
 \ssr 84, 297, V-M98

\end{references}
\end{document}